# A New Hope for Long Space Flights: Hypercapnia demonstrated suppression of metabolic acidosis in an experiment with prolonged stay of people in a closed life support system

Semyonov D.A., Semyonova A.A.

Prof. V.F. Voino-Yasenetsky Krasnoyarsk State Medical University, Krasnoyarsk, Russia

**Introduction.**

Sixty-four years have passed since the first human space flight. The record in terms of duration of human space flight was realized 30 years ago. For all this time humanity has not been able to solve many problems that threaten human health in a long space flight. An unexpected solution is suggested by an experiment on human stay in a model of a closed life support system, conducted more than 50 years ago in Krasnoyarsk. The high but compatible with active life activity level of carbon dioxide, which was established after the beginning of the experiment, changed the metabolism of the subjects. Although such changes in the human body under the influence of hypercapnia have been studied many times over the past half century, but the interpretation of the results of the Krasnoyarsk experiment through understanding the effects of hypercapnia on metabolism we present for the first time in this paper. Hypercapnia suppressed the possibility of metabolic acidosis and activated fat oxidation. The hypercapnia-induced change in respiratory quotient stabilized the gas mixture composition in the closed-loop life support system and ensured system stability. Suppression of metabolic acidosis by hypercapnia may have several positive implications for manned space missions, in particular, there is hope to combat calcium loss by bones. We propose a project to incorporate regular personalized hypercapnic treatments during long-duration space missions.

1.  **Metabolic acidosis,** i.e. an increase in plasma lactate concentration above 2mM, is a common condition associated with acute and chronic diseases (Nechipurenko et al., 2021). Metabolic acidosis is also called lactate acidosis, in the text we will use the terms as equivalent.

In sports biochemistry, lactate levels near 2mM have empirically been recognized as an important threshold level (metabolic threshold).

The inhibitory effect of lactate on free fatty acid concentrations in mammalian plasma was shown experimentally as early as the early 1960s (Issekutz and Miller, 1962).

The first works predicting a switch to carbohydrate oxidation during athletic exertion also appeared in the 60s. Initially, Wasserman's works (Wasserman & McIlroy, 1964; Wasserman et al., 1967) suggested that there should be a single threshold after which hypoxia, the inability to oxidize fats efficiently, an increase in lactate concentration, and a significant increase in the release of carbon dioxide during the absorption of one liter of oxygen develop in concert. However, in reality, several threshold phenomena have been found to be registered at the level of the whole organism.

In tests with gradually incremental exercise, the dynamics of lactate concentration growth in capillary blood changes qualitatively near the concentration of 2mM. The graph of increasing equilibrium lactate concentration up to some load is a slowly growing, gentle, almost horizontal line, well approximated by a straight line, and afterwards there is an active nonlinear increase in lactate concentration. This threshold attracts the most attention in the development of endurance athletes, that is, in sports such as running, cycling, and long-distance swimming. Details of the measurement methodology can be found in (Goodwin et al., 2007), where the moment of increase in lactate concentration is referred to simply as the "lactate threshold". We will also use the commonly used term aerobic threshold. Why does this threshold exist and even claims to be universal at the level of the whole organism? The point is that a trained organism in response to a moderate-intensity load can afford not to panic and not to rush to switch from energy-dense fats to carbohydrates. A trained body has an adequate supply of oxygen to its muscles. In response to a moderate load, which the body perceives as light, an additional amount of fat is mobilized, i.e. lipolysis is activated in the cells of adipose tissue. As the load increases, the amount of oxidized fat initially increases as well. At the same time, the body gradually moves to more and more oxidation of carbohydrates, the level of lactate in the blood gradually rises. But as soon as the absolute amount of oxidized fats passes the maximum, this flow is forced to be replaced by carbohydrates, which leads to an acceleration of lactate production. Since 2001, there has been the idea of optimizing exercise near the maximum of fat oxidation (Jeukendrup, Achten, 2001) and (Achten, Jeukendrup, 2004). It is the passage through this maximum that provokes the first spike in lactate concentration, and this is how the existence of the aerobic threshold can be interpreted (Bircher at al., 2005).

Lactate is able to inhibit lipolysis in adipocytes (Brooks, 2020). That is, lactate acts as a hormone that regulates fat mobilization. Initially, it is the loaded muscle that has an increased need for fat. If the flow of incoming fatty acids and oxygen do not compensate for the muscle's energy demand, the muscle sends a signal to switch to a different energy supply mode. It is in the muscle with increasing load that moderate hypoxia develops and the need to switch to carbohydrate oxidation appears, but the signal that lipolysis can be reduced to a basic level must penetrate into the adipose tissue precisely by means of capillary blood. The signal must be triggered as quickly as possible, i.e. there is a need for regulation via lactate receptors on the cell surface. HCAR-1 (or GPR81 which is the same thing) has been shown to be such a receptor (Ahmed et al, 2010). The regulation of this process through a single receptor on the surface of adipocytes, contacting almost directly with blood plasma makes the concentration value of 2mM in capillary blood almost universal. The HCAR-1 receptor is not only found on the surface of adipocytes and muscle cells (Brooks, 2020). The receptor is present on the surface of endothelial cells (Sun et al., 2019). And it is also present on the surface of immune cells, particularly macrophages and dendritic cells (Manoharan et al., 2021). It is likely that in all these cases, the overcoming of aerobic threshold by the organism may be a significant event for cells that have such a receptor on their surface (Figure1).

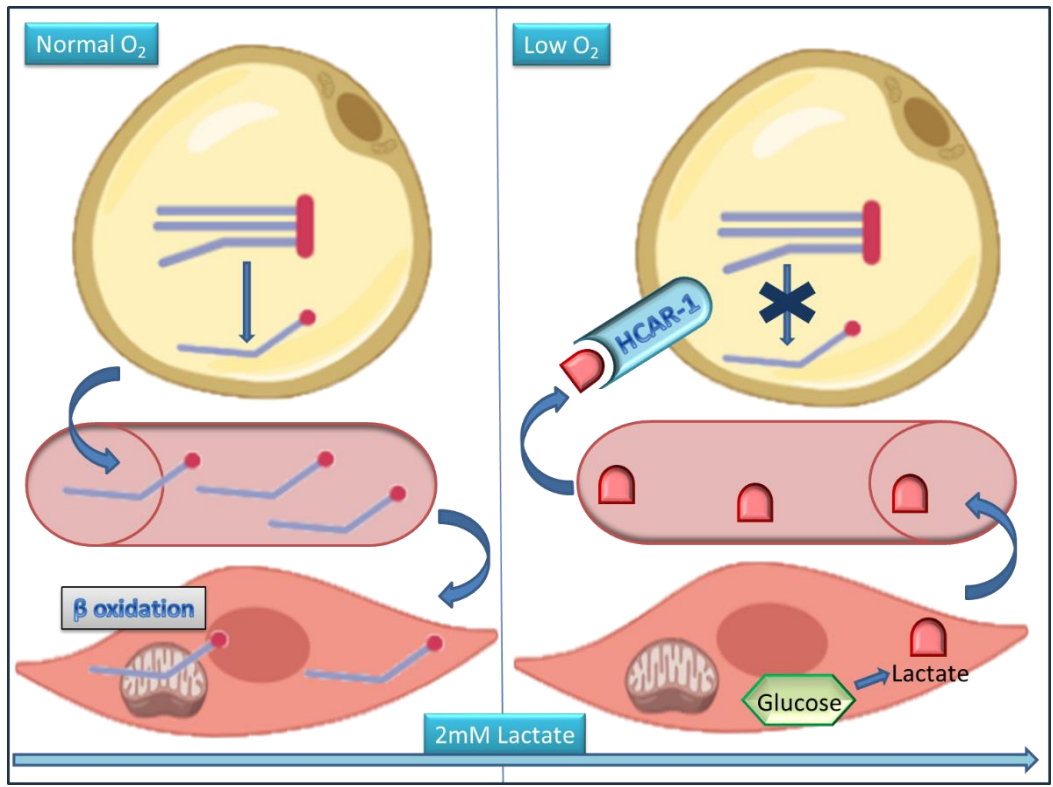

Figure 1. Effect of lactate concentration on lipolysis in adipocytes. The threshold value of lactate 2 mM qualitatively changes metabolism: lipolysis is inhibited, glycolysis is activated.

Metabolic acidosis was observed throughout the entire time of astronaut Scott Kelly's stay on the ISS (Garrett-Bakelman et al., 2019). Judging by the publications, the issue of measuring lactate concentration in astronauts in orbit has not interested researchers for a long time. Even after a major program to develop biochemical diagnostic methods for space station applications, lactate did not become popular. American astronauts used measuring lactate concentrations after exercise to monitor the body's response to exertion, but resting lactate measurements do not appear to have been performed prior to Scott Kelly's flight. Training of American astronauts does not result in a decrease in muscle mass in weightlessness, which is perceived as a significant success. The focus on maintaining muscle mass is similar to the American approach to something like CrossFit: short daily strength and speed training. Metabolic acidosis at rest and high post-exercise lactate concentrations do not prevent CrossFit athletes from maintaining muscle mass, but they do significantly reduce cognitive performance (Perciavalle et al., 2016).

The training of cosmonaut Valery Poliakov, the world record holder in orbit, was not accompanied by daily lactate measurements at rest, as far as we know (Poliakov, Noskov, 2005). Nevertheless, the system of long-term low-intensity training developed by V.V. Polyakov's can be considered as a way to prevent metabolic acidosis. Although the Russian approach to training in orbit has now changed, moving closer to the American approach (Fomina et al., 2016, Trappe et al., 2009), Polyakov's training has proven to be effective in maintaining the cosmonaut's metabolic health. Polyakov trained daily for up to 6.5 hours at low intensity, which is exactly the same as the training regimen of cyclists or triathletes, where training aims to increase mitochondrial numbers and intensify fat oxidation (Faria et al., 2005). A similar approach exists in other endurance sports (Casado et al., 2023). Polyakov's approach simultaneously demonstrated biological success and economic futility: 6.5 hours of training is a full day's work, i.e., the cosmonaut can no longer fully perform other work while maintaining his or her own health. Polyakov's success was at the same time a demonstration of the crisis of manned cosmonautics, which has not been overcome to this day.

**2. Possible causes and mechanisms of metabolic acidosis provocation in space flight and in lunar base conditions.** *Disturbance of hemodynamics*. In terrestrial conditions under the action of gravity venous blood with difficulty returns to the heart from the lower extremities. The local hypoxia created in the leg muscles can provoke the production of additional lactate. Moderate exercise of the leg muscles improves blood circulation and also allows the utilization of lactate synthesized in the muscles. In weightlessness, blood does not accumulate in the legs, but in the chest and face (Hussain et al., 2024). Because of the lack of natural ways to overcome this redistribution of blood, the local hypoxia that develops is more difficult to overcome with leg exercises. *The hypodynamia* that develops in weightlessness is caused not only by insufficient exercise of the leg muscles, but also by the non-use of many postural muscles involved in maintaining posture (most often in stabilizing the vertical position of the body). Such muscles are usually rich in red muscle fibers that actively oxidize fat in mitochondria and produce low lactate. Prolonged hypodynamia provokes a decrease in fat requirement and a switch to glucose oxidation. That is, an additional source of lactate appears. *Stress*, acting both short-term and long-term, leads to activation of lactate synthesis in muscles. Under the action of adrenaline and cortisol in muscle cells, glycogen is broken down to form glucose-6-phosphate. There is almost no enzyme in muscles that cleaves phosphate from glucose, so the further chain of transformations leads to the formation of first pyruvate and then lactate. Pyruvate could be spent by muscle work, but in weightlessness muscles are not involved. As a result of stress on the human body, muscles in weightlessness become sources of lactate. In future interplanetary

missions and during base construction, for example on the Moon, another factor triggering lactate synthesis will be a *carbohydrate-rich diet*. Optimization of the diet of participants in future long-duration space missions should take into account the fact that excess carbohydrates in the diet can provoke metabolic acidosis. (Figure 2) Low-intensity physical activity can compensate for all of these negative factors, but the amount of exercise required is high. The demands on the fitness level of the mission participants are also great. It would be desirable that the opportunity to participate in such missions is not limited to the cycling and triathlon community. In addition, during sleep, factors that provoke metabolic acidosis continue to act on the human body. Finding an alternative to prolonged low-intensity exercise appears to be relevant to preventing metabolic acidosis.

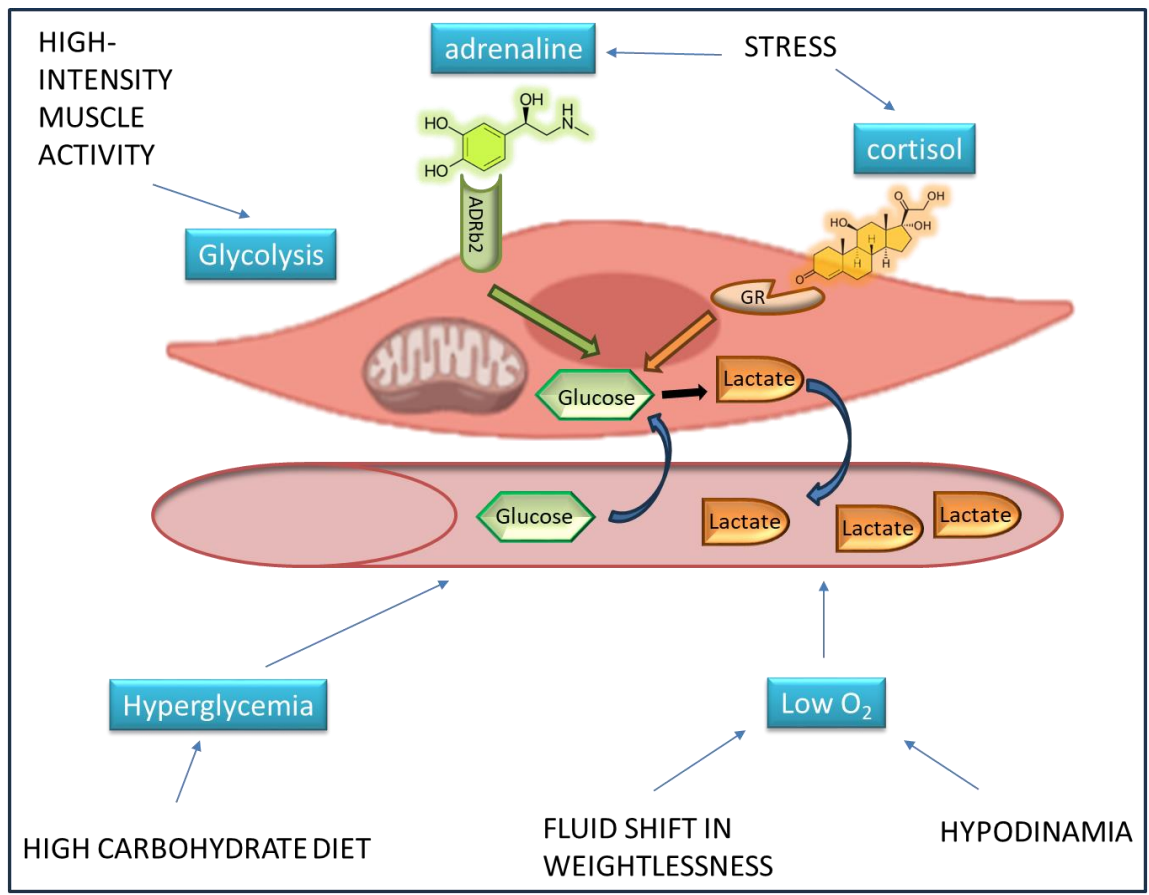

Figure 2: The reasons contributing to the development of metabolic acidosis.

**3. Is chronic metabolic acidosis so bad?** As mentioned above, prolonged elevation of lactate levels is a sign of hypoxia and causes adaptive changes in the hematopoietic system. In addition, metabolic acidosis is perceived as a consequence of many diseases. The modern view of lactate, however, makes it necessary to consider metabolic acidosis not as a consequence, but as a metabolic background on which a chronic disease develops. For example, the turn off of lipolysis when lactate levels rise is the metabolic background, and possibly the cause, for the development of obesity and type 2 diabetes (Ahmed et al, 2010). Since the discovery of the Warburg effect, lactate has been linked to oncogenesis (Warburg, 1956; Gatenby,

Gillies, 2004). At present, it can be noted that the notion of metabolic acidosis as a metabolic background for oncogenesis seems to be increasingly valid (San-Millan, Brooks, 2017). More recently, lactate has been announced as a key molecule explaining the mechanism of cachexia development in cancer (Liu et al., 2024). Macrophages have receptors for lactate on their surface (Manoharan et al., 2021), which can probably be related to their ability to metabolically reprogram and switch from a resting state when they oxidize fats to an inflammatory phenotype that oxidizes glucose (Larionova et al., 2020). Receptors to lactate are also present on the endothelium of blood vessels, which may be related to the regulation of their tone and blood flow under hypoxia (Jones et al., 2020; Wu et al., 2023. It is conceivable that such regulation, which is necessary for adaptation to exercise, may be impaired by prolonged metabolic acidosis. There are receptors for lactate in the brain as well (Mosienko et al., 2015), indicating a possible role of lactate in the regulation of higher nervous activity and behavior. More recently, lactate has been discovered to be directly involved in the regulation of gene expression through the mechanism of histone lactylation (Zhang et al., 2019). The diversity and scope of regulation in which lactate is involved is remarkable, as reflected in current reviews on its role in biochemistry and physiology (Brooks, 2020). (Figure 3)

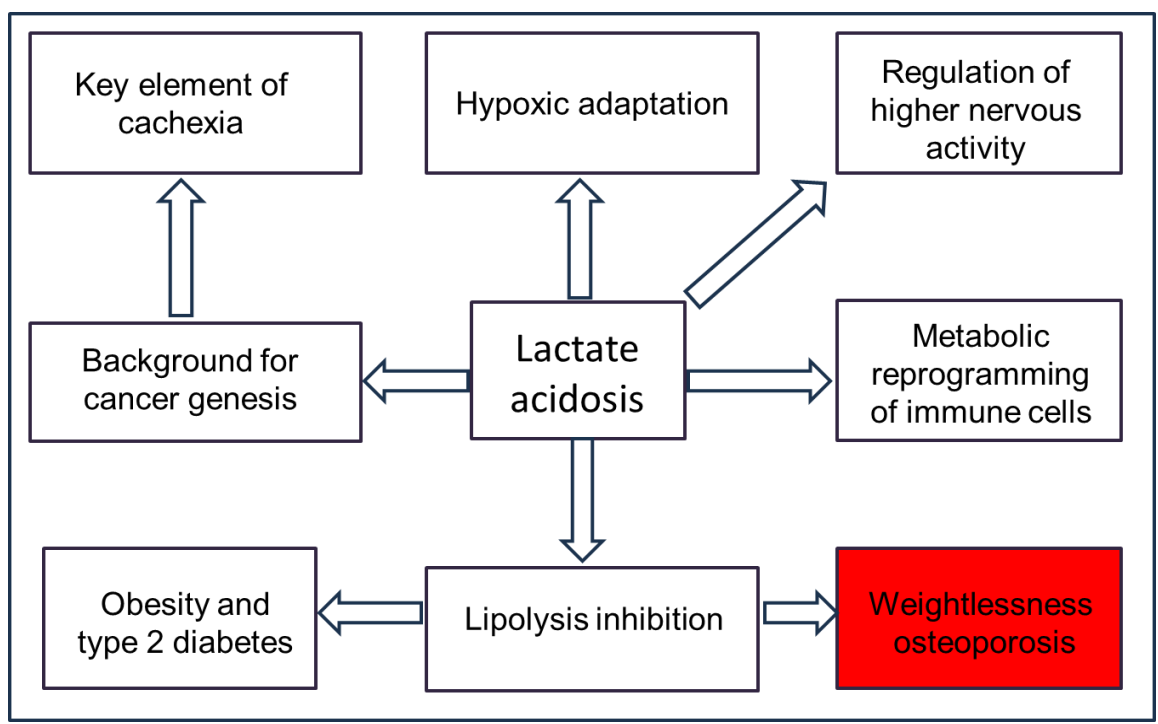

Figure 3: Implications of a metabolic background of lactate.

A metabolically healthy person should not be in a state of metabolic acidosis. There is a syndrome that develops specifically in weightlessness, and possibly in reduced gravity, that is probably closely related to metabolic acidosis. Let us focus in detail on such a phenomenon as the loss of calcium by the human body in conditions of reduced gravity. In the recent work of Canadian scientists demonstrated that during the stay in orbit in the bones of astronauts deposited fat (Liu et al., 2023). Upon return to Earth, by oxidizing these very fat stores in the bone marrow, the body gains energy to partially restore bone mineral matter. Normal Earth patients with osteoporosis have also shown an increase in bone fat stores (Yeung et al, 2005). Compare this to the story of Scott Kelly's development of metabolic acidosis and the recovery of his healthy metabolism after returning to Earth. The metabolic acidosis developed in orbit blocked lipolysis and triggered lipogenesis. As a result, bone fat stores increased and the ability to obtain energy for bone mineralization was lost. In addition to this, lactate may well leach calcium from the bones, increasing the rate at which the body loses calcium. This view of osteoporosis in astronauts differs significantly from the notion that calcium leaves the bones in weightlessness simply because bone strength is no longer needed.

First and foremost, this leads one to focus on the possibility of preventing metabolic acidosis.

## 4. Hypercapnia and metabolism. Respiratory acidosis versus metabolic acidosis.

Hypercapnia is an effect that can lower the plasma pH of alveolar capillaries (respiratory acidosis). Excess lactic acid (metabolic acidosis) in plasma is also an effect that can lower the pH. As a result of the development of respiratory acidosis under the influence of hypercapnia, the human body, which retains metabolic plasticity, is able to reduce the level of lactic acid in the blood plasma. That is, respiratory acidosis is able to suppress metabolic acidosis. If the costs of the required exercise time seem too great, the use of hypercapnia could reduce the duration or even replace prolonged moderate-intensity exercise altogether.

Let us briefly consider the physiological or even biophysical underlying causes of this phenomenon.

The local oxygen concentration required to maintain normal cell metabolism in peripheral tissues is determined by the pH in these tissues. When intense metabolism leads to a decrease in pH, hemoglobin more easily gives up $O_2$ , while binding excess protons, thereby providing efficient transport of oxygen from the lungs to the tissues, and transport of carbon dioxide (mainly in the form of bicarbonate) in the opposite direction. This negative feedback metabolite regulation system based on a cooperative pH-dependent change in hemoglobin conformation is known as the Bohr effect (Bohr et al., 1904, Ahmed et al., 2020)

A slight deviation of blood acidity from the physiological norm can significantly alter the ability of hemoglobin to bind oxygen. It should be emphasized that the Bohr effect is a key element in the regulation of gas exchange in humans and many animals. A decrease in blood plasma pH from 7.4 to 7.2 leads to a twofold decrease in the amount of O2 that can bind hemoglobin at a partial pressure of oxygen in tissue fluid of the order of 20-40 mm Hg. Therefore, during oxygenation, it is critical for the body to maintain an optimal pH value in the blood plasma of the alveolar capillaries.

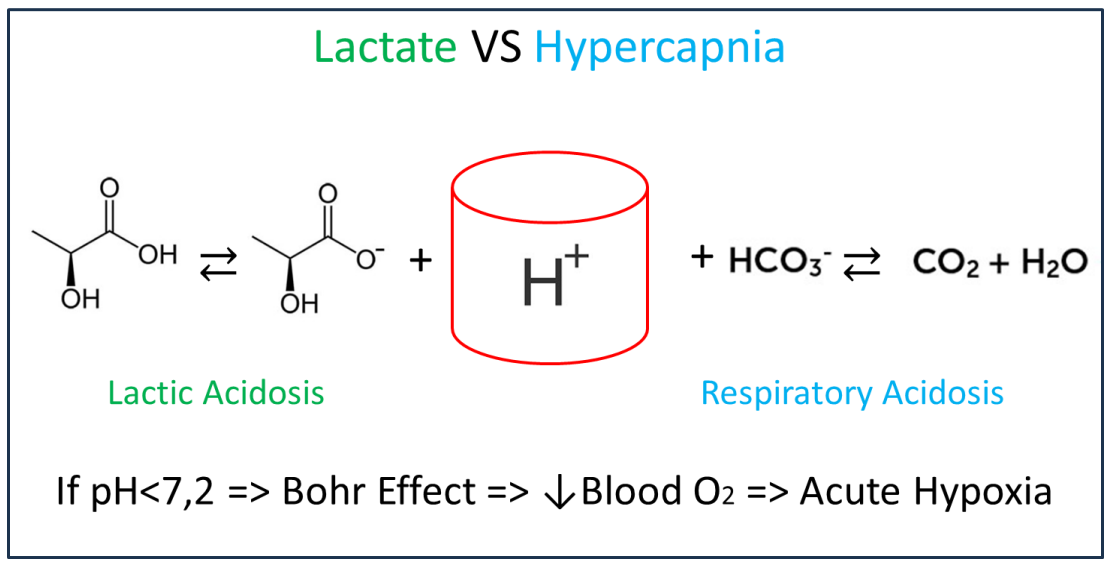

Figure 4: Opposition of lactate and respiratory acidosis.

Lung plasma acidity, that is, the concentration of hydrogen ions, is generated from two major sources: dissociation of lactic acid and dissolution of carbon dioxide followed by dissociation to protons and hydrogen carbonate. The two sources fill a single volume with hydrogen ions (Figure 4). And in the case of buffer overflow, we face the unfortunate consequences so well known to many from the recent COVID-19 pandemic: a drop in blood saturation and oxygen deficiency. It is natural to assume that the body has a mechanism to actively utilize lactate to prevent the development of acidosis to a pH that threatens a drop in saturation. Lung tissue itself, containing a mass of mitochondria and lactate dehydrogenase-III isoenzyme, could actively remove lactate from blood plasma (Drent et al., 1996). Although the molecular mechanisms of lactate removal from blood plasma during hypercapnia are not yet fully understood, the evidence of the existence of the phenomenon itself has been obtained to a sufficient extent and for quite a long time. Thus, in a clinical study back in 1976, a significant decrease in plasma lactate concentration in all athletes who received a significant exercise (70% of the maximum) with simultaneous inhalation of air with 4% CO2 was recorded (Rizzo et al., 1976). Slightly later, but also in athletes at exercise, a decrease in lactate concentration was demonstrated with a stepwise increase in inhaled air CO2 concentration from 0 to 6% (Graham et al., 1982). The authors suggested that hypercapnia suppresses carbohydrate oxidation and activates fatty acid oxidation. Two clinical studies in 2005 confirmed that hypercapnia reduced lactate production (Kato et al., 2005; Li et al., 2005) in both athletes and children in the intensive care unit. Experience with permissive hypercapnia also indicates suppression of metabolic acidosis (Hickling, Joyce, 1995)

Of course, a critical excess of carbon dioxide as well as hyperproduction of lactate by the body will lead to acidosis and a subsequent drop in saturation. It is important for us to emphasize, however, that moderate hypercapnia can suppress moderate metabolic acidosis. In intensive care practice, when the patient's condition is monitored by blood gas and electrolyte analysis, the opposition between respiratory and metabolic acidosis can be seen clearly: changes in respiratory acidosis are similar to metabolic alkalosis and, conversely, respiratory alkalosis leads to metabolic changes similar to metabolic acidosis (Hyneck, 1985).

## 5. Prolonged Hypercapnia in the BIOS-3 Experiment

Can astronauts withstand prolonged exposure to hypercapnia? Will prolonged hypercapnia result in the expected effect of suppressing metabolic acidosis? An experiment on human exposure to prolonged continuous hypercapnia was conducted more than 50 years ago. This experiment was directly related to the Soviet space program, but the result we are interested in was not understood by the researchers.

More than 50 years ago, Krasnoyarsk biophysicists created the BIOS-3 system modeling a closed ecosystem for human life support. The longest experiment with human participation (180 days) was also conducted half a century ago. The prerequisites for the creation of BIOS-3 and the results of the experiments are described in (Gitelzon et al., 1975), where data are also given that the test subjects were kept under conditions of moderate hypercapnia for a long time. The creators of BIOS-3 initially assumed that the increased concentration of carbon dioxide would not be a significant factor and interpreted all further results in this way. Moreover, having postulated that the key link in the life support system is the human body, the creators of BIOS-3 actually made plants the key link. On the basis of literature data alone, they tried to justify the claim that a carbon dioxide concentration of 1% will be safe for any length of time. The main reason for choosing such a concentration of carbon dioxide can be considered to be its effect on plant growth (Gitelson et al., 1975). The experiment conducted 50 years ago was feasible and successful due to the metabolic plasticity of the human body. To deny this would mean to voluntarily give up the opportunity to make useful conclusions both for manned cosmonautics and for designing life support systems for other needs.

It should be noted that the creators of BIOS-3 were aware in advance of the possible difficulties in adapting to carbon dioxide concentrations near 1%. The literature they analyzed at the time indicated that such a carbon dioxide concentration was not readily accepted. Nevertheless, some tentative optimism was maintained through the belief that adaptation within three weeks was possible. It should be noted that this or a close characteristic time coincides with the adaptation time to mid-mountain conditions. This may be related to the reorganization of the erythropoiesis system, including erythropoietin production. Stimulation of erythropoiesis allows to compensate hypoxia. For athletes-skiers staying in the middle mountains opens up opportunities to achieve higher results, and for mountaineers there are opportunities to climb higher in the mountains without severe health consequences. How did the erythropoiesis of BIOS-3 inhabitants react? All observations of BIOS-3 test subjects indicate no significant increase in hemoglobin concentration and adaptation to hypoxia. At the same time, there were episodes of decreased atmospheric oxygen levels in the four-month experiment (Terskov et al., 1979). The lack of adaptation can be explained by the presumed absence of increased lactate concentration, since glycolysis was suppressed by constant hypercapnia. Probably, metabolic acidosis in BIOS-3 testers simply did not develop, which is why their metabolism differed not only from skiers and mountaineers, but also from astronauts.

The effect of metabolic acidosis on the regulation of erythropoiesis and erythropoietin production was actively studied at the same time (Rodgers et al, 1974; Rodgers et al, 1975). According to the ideas of the time, lactate was supposed to be responsible for the stimulation of erythropoiesis during hypoxia. Although the breathing of the testers became more rapid when the atmosphere in the system contained an increased concentration of carbon dioxide and a slightly reduced concentration of oxygen (about 19%), but the hemoglobin level in the blood and blood pressure changed insignificantly (Gitelzon et al., 1975).

Accounting for the effect of hypercapnia on the metabolism of the testers is also important because humans were the main link releasing carbon dioxide and returning it back to the atmosphere. That is, not only did carbon dioxide affect human metabolism, but human metabolism affected the equilibrium concentration of carbon dioxide in the closed system. If lactic acid production is depressed as carbon dioxide concentration increases, this is equivalent to a decrease in the proportion of carbohydrates in metabolism. A decrease in the rate of carbohydrate oxidation should be compensated by an increase in the rate of fat oxidation. This is exactly what is found in the data published by the creators of BIOS-3 in the journal Voprosy Pitaniya (suitable translation of the journal name: Nutrition Issues) back in 1976 (Vlasova et al., 1976). The level of triglycerides in blood plasma under hypercapnia conditions was significantly elevated in only one test subject out of four. The authors of the 1976 paper and the authors of the BIOS-3 experiment did not in any way relate the significant increase in the concentration of lipids in the blood of the tester to the synchronous increase in hypercapnia and the subsequent decline in fat intake to the normalization of atmospheric composition. A more remarkable fact is that the level of free fatty acids in the blood of all subjects during the experimental period was at the same level as before the experiment. That is, metabolic acidosis did not develop.

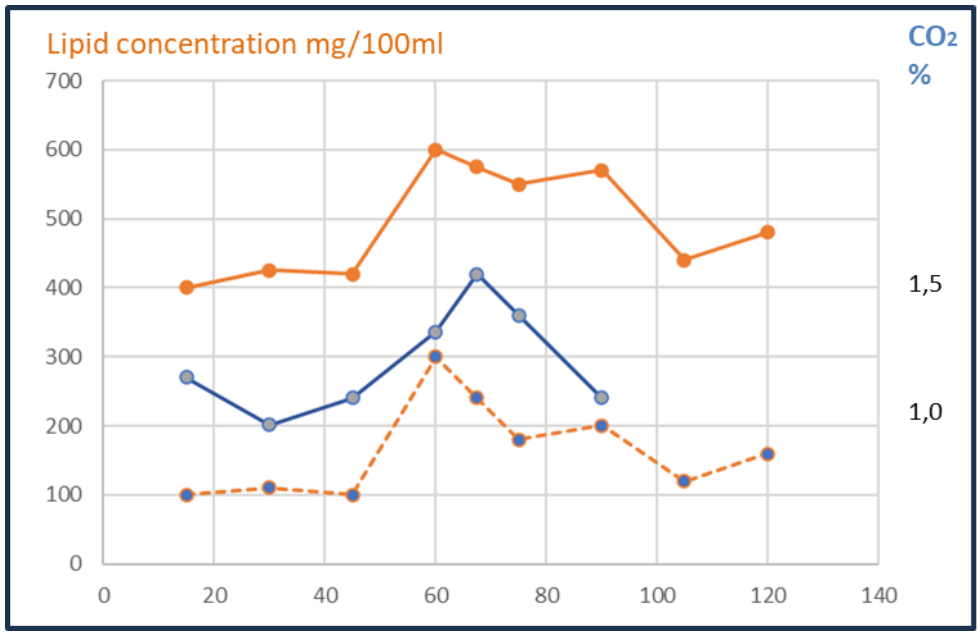

Figure 5. Dynamics of blood lipid concentrations in one of the BIOS-3 unit testers and % carbon dioxide in the unit. Orange curve - total lipids, dotted curve - triglycerides, blue curve - carbon dioxide. X-axis - time, days

During this period, the metabolic transition from fats to carbohydrates and back was already experimentally studied, but in relation to the physiology and biochemistry of physical exercise. In conditions of approaching hypoxia the organism switches to oxidation of carbohydrates, lactate production increases, lactate blocks lipolysis through receptors on adipocytes. Normal and even slightly increased levels of free fatty acids directly indicates that the test subjects' bodies did not feel hypoxia and did not undergo metabolic acidosis.

Note that suppression of lactate production leads to a decrease in the respiratory coefficient, i.e. to a decrease in carbon dioxide production while absorbing the same amount of oxygen. Under conditions of a closed system, a negative feedback is formed: an increase in carbon dioxide concentration leads to a decrease in carbon dioxide production by the human body. That is, in the BIOS-3 experiment, the equilibrium composition of the gas environment was determined by the ability of hypercapnia to influence metabolism. If there was no this influence, a stable equilibrium state would not occur in a closed life-support system. And the development of acidosis would have destabilized the gas composition of the closed life support system. This equilibrium was not predicted in advance when designing the life support system, but fortunately, the BIOS-3 project was successful. It is interesting to realize the possibility that the key link in stabilizing the gas composition of the closed system in the experiment turned out to be the test subject with the greatest manifestations of lipid metabolism activation. The much later Biosphere-2 project (USA) ended with the evacuation of the testers due to a drop in the oxygen level and an increase in the carbon dioxide level to values incompatible with human life.

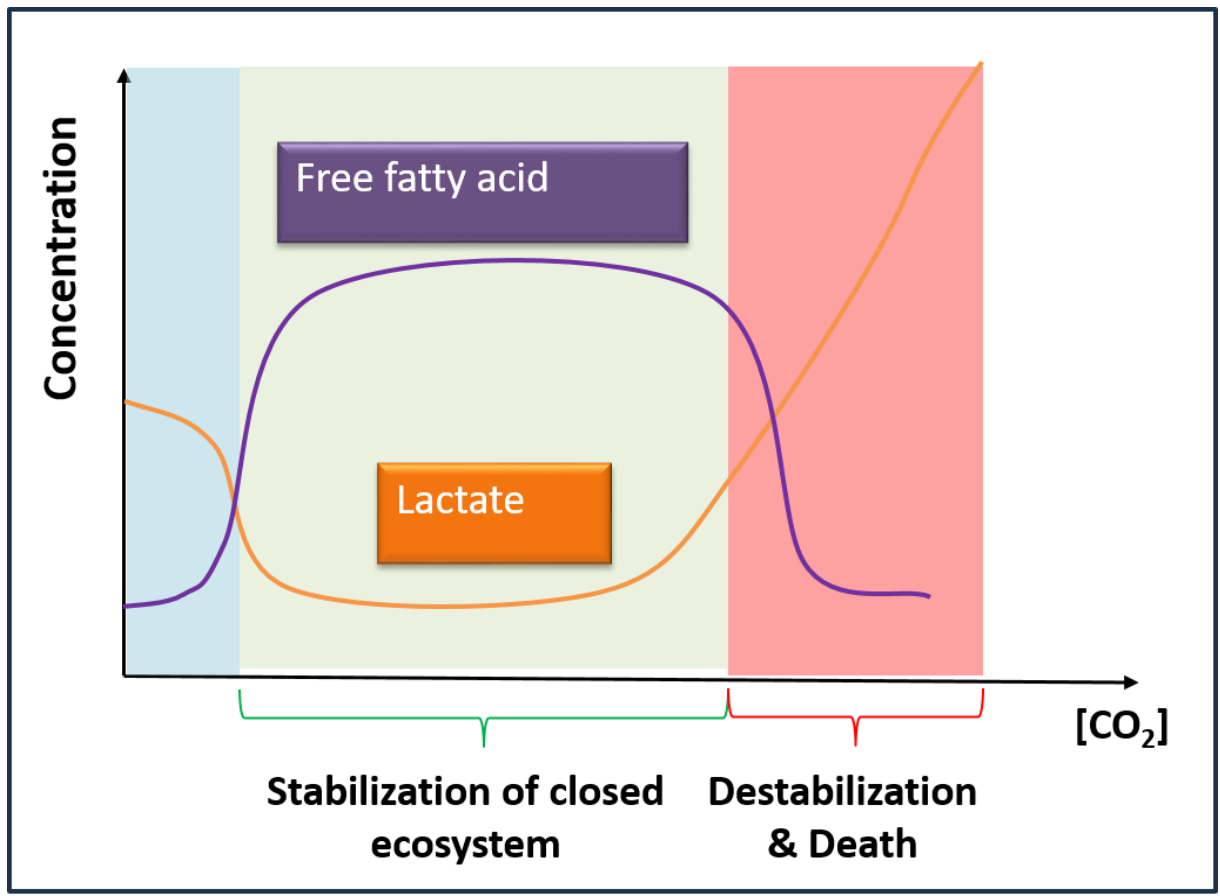

Figure 6: Change in the concentration of free fatty acids and lactate depending on the concentration of carbon dioxide

**Conclusion.**

Metabolic acidosis in the organism of the inhabitants of a closed life support system is not only harmful to health, but also dangerous for the whole life support system. The method of maintaining metabolic health through regular long-term low-intensity training found and successfully tested by Valery Polyakov is time-consuming. We propose to replace Polyakov's approach with regular controlled hypercapnia procedures. We propose that the procedure be personalized. That is, hypercapnia can be included as a procedure to be performed several times a day. The use of personalized equipment would allow the habitual composition of the inhaled air to be maintained between procedures. Each crew member would receive sufficient exposure to inhibit metabolic acidosis and activate lipolysis. Each participant would also make a personalized contribution to maintaining the desired respiratory quotient.

Respiratory acidosis vs. metabolic acidosis: hypercapnia causes a decrease in lactate

Polyakov's method: prolonged exercise of moderate intensity causes a decrease in lactate

The BIOS-3 experiment did not give an answer about the exhaustibility of the resource of hypercapnia application. Three months after the beginning of the experiment, the conditions had changed significantly. Continuous hypercapnic exposure may be significantly different from intermittent exposure. Prolonged and ultra-long-term use of hypercapnia may differ in its results from short-term procedures in a research laboratory or ICU setting. The ultimate understanding of the prospects for the use of hypercapnia to normalize the metabolism of participants in long-duration space missions depends on further experimental studies.

We are not suggesting that exercise and the selection of a rational diet should be abandoned. Moreover, we believe that the interaction of hypercapnia, exercise and diet should be studied in a serious way. Clearly, a trained individual will respond differently to hypercapnia than an untrained individual. A potentially untrained and weakened individual may respond by immediately developing acute hypoxia.

We also believe that we need to pay careful attention not only to the training of participants in long-duration space missions, but also to the selection of crew candidates. Perhaps some candidates will initially be genetically better adapted to prolonged exposure of their metabolism to external factors. It is now no surprise to anyone that Kenyan Kalenji runners are better adapted to the marathon and Nepalese Gurkhas are adapted to high altitudes. The diversity of human genotypes gives mankind a chance to cope with supernatural challenges such as long-duration space missions.

Besides the prospect of overcoming the crisis in health and life support for participants in manned space missions, hypercapnia may help in terrestrial medicine. Metabolic acidosis is the background for the development of many pathologies. In addition to weightlessness osteoporosis, metabolic acidosis is also present in normal osteoporosis. It was also mentioned in this text that metabolic acidosis is a background for the development of obesity and type 2 diabetes, cancer and hematopoietic disorders. It is likely that understanding of the role of lactate in disease development will increase as this issue is studied. We would not rule out the involvement of lactate in autoimmune, neurodegenerative and age-related diseases. Perhaps the technology developed to send humans on a long-duration space mission will then be needed on Earth as much as in Space.

Literature:

Achten J, Jeukendrup AE. Relation between plasma lactate concentration and fat oxidation rates over a wide range of exercise intensities. Int J Sports Med. 2004 Jan;25(1):32-7. doi: 10.1055/s-2003-45231. PMID: 14750010.

Ahmed K, Tunaru S, Tang C, Müller M, Gille A, Sassmann A, Hanson J, Offermanns S. An autocrine lactate loop mediates insulin-dependent inhibition of lipolysis through GPR81. Cell Metab. 2010 Apr 7;11(4):311-9. doi: 10.1016/j.cmet.2010.02.012. PMID: 20374963.

Ahmed, M.H.; Ghatge, M.S.; Safo, M.K. Hemoglobin: Structure, function and allostery. Subcell. Biochem. 2020, 94, 345–382.

Bircher S, Knechtle B, Knecht H. Is the intensity of the highest fat oxidation at the lactate concentration of 2 mmol L(-1)? A comparison of two different exercise protocols. Eur J Clin Invest. 2005 Aug;35(8):491-8. doi: 10.1111/j.1365-2362.2005.01538.x. PMID: 16101669.

Bohr, C.; Hasselbalch, K.; Krogh, A. Concerning a biologically important relationship–the influence of the carbon dioxide content of blood on its oxygen binding. Skand. Arch. Physiol. 1904, 16, 401–412.

Brooks GA. Lactate as a fulcrum of metabolism. Redox Biol. 2020 Aug;35:101454. doi: 10.1016/j.redox.2020.101454. Epub 2020 Feb 9. PMID: 32113910; PMCID: PMC7284908.

Casado A, Foster C, Bakken M, Tjelta LI. Does Lactate-Guided Threshold Interval Training within a High-Volume Low-Intensity Approach Represent the "Next Step" in the Evolution of Distance Running Training? *International Journal of Environmental Research and Public Health*. 2023; 20(5):3782. https://doi.org/10.3390/ijerph20053782.

Drent M, Cobben NA, Henderson RF, Wouters EF, van Dieijen-Visser M. Usefulness of lactate dehydrogenase and its isoenzymes as indicators of lung damage or inflammation. Eur Respir J. 1996 Aug;9(8):1736-42. doi: 10.1183/09031936.96.09081736. PMID: 8866602.

Faria EW, Parker DL, Faria IE. The science of cycling: physiology and training - part 1. Sports Med. 2005;35(4):285-312. doi: 10.2165/00007256-200535040-00002. PMID: 15831059.

Fomina EV, Lysova NY, Chernova MV, Khustnudinova DR, Kozlovskaya IB. [Comparative Analisys of Efficacy of Countermeasure Provided by Different Modes of Locomotor Training in Space Flight.]. Fiziol Cheloveka. 2016 Sep;42(5):84-91. Russian. PMID: 29932552.

Garrett-Bakelman FE, Darshi M, Green SJ, Gur RC, Lin L, Macias BR, McKenna MJ, Meydan C, Mishra T, Nasrini J, Piening BD, Rizzardi LF, Sharma K, Siamwala JH, Taylor L, Vitaterna MH, Afkarian M, Afshinnekoo E, Ahadi S, Ambati A, Arya M, Bezdan D, Callahan CM, Chen S, Choi AMK, Chlipala GE, Contrepois K, Covington M, Crucian BE, De Vivo I, Dinges DF, Ebert DJ, Feinberg JI, Gandara JA, George KA, Goutsias J, Grills GS, Hargens AR, Heer M, Hillary RP, Hoofnagle AN, Hook VYH, Jenkinson G, Jiang P, Keshavarzian A, Laurie SS, Lee-McMullen B, Lumpkins SB, MacKay M, Maienschein-Cline MG, Melnick AM, Moore TM, Nakahira K, Patel HH, Pietrzyk R, Rao V, Saito R, Salins DN, Schilling JM, Sears DD, Sheridan CK, Stenger MB, Tryggvadottir R, Urban AE, Vaisar T, Van Espen B, Zhang J, Ziegler MG, Zwart SR, Charles JB, Kundrot CE, Scott GBI, Bailey SM, Basner M, Feinberg AP, Lee SMC, Mason CE, Mignot E, Rana BK, Smith SM, Snyder MP, Turek FW. The NASA Twins Study: A multidimensional analysis of a year-long human spaceflight. Science. 2019 Apr 12;364(6436):eaau8650. doi: 10.1126/science.aau8650. PMID: 30975860; PMCID: PMC7580864.

Gatenby RA, Gillies RJ. Why do cancers have high aerobic glycolysis? Nat Rev Cancer. 2004 Nov;4(11):891-9. doi: 10.1038/nrc1478. PMID: 15516961.

Gitelzon I.I., Kovrov B.G., Lisovsky G.M., et al. Experimental ecological systems including humans. Problems of Space Biology vol. 28. Moscow, - Nauka, 1975, 312 p. Russian.

Goodwin ML, Harris JE, Hernández A, Gladden LB. Blood lactate measurements and analysis during exercise: a guide for clinicians. J Diabetes Sci Technol. 2007 Jul;1(4):558-69. doi: 10.1177/193229680700100414. PMID: 19885119; PMCID: PMC2769631.

Graham TE, Wilson BA, Sample M, Van Dijk J, Goslin B. The effects of hypercapnia on the metabolic response to steady-state exercise. Med Sci Sports Exerc. 1982;14(4):286-91. doi: 10.1249/00005768-198204000-00006. PMID: 7132646.

Hickling KG, Joyce C. Permissive hypercapnia in ARDS and its effect on tissue oxygenation. Acta Anaesthesiol Scand Suppl. 1995;107:201-8. doi: 10.1111/j.1399-6576.1995.tb04359.x. PMID: 8599278.

Hussain I, Ullah R, Simran BFNU, Kaur P, Kumar M, Raj R, Faraz M, Mehmoodi A, Malik J. Cardiovascular effects of long-duration space flight. Health Sci Rep. 2024 Aug 12;7(8):e2305. doi: 10.1002/hsr2.2305. PMID: 39135704; PMCID: PMC11318032.

Hyneck ML. Simple acid-base disorders. Am J Hosp Pharm. 1985 Sep;42(9):1992-2004. PMID: 3931455.

Issekutz, B. v. and Harvey I. Miller. "Plasma Free Fatty Acids During Exercise and the Effect of Lactic Acid." *Proceedings of the Society for Experimental Biology and Medicine* 110 (1962): 237 - 239.

Jeukendrup, A., Achten, J. Fatmax: A new concept to optimize fat oxidation during exercise?. European Journal of Sport Science, 2001, 1: 1-5. https://doi.org/10.1080/17461390100071507

Jones NK, Stewart K, Czopek A, Menzies RI, Thomson A, Moran CM, Cairns C, Conway BR, Denby L, Livingstone DEW, Wiseman J, Hadoke PW, Webb DJ, Dhaun N, Dear JW, Mullins JJ, Bailey MA. Endothelin-1 Mediates the Systemic and Renal Hemodynamic Effects of GPR81 Activation. Hypertension. 2020 May;75(5):1213-1222. doi: 10.1161/HYPERTENSIONAHA.119.14308. Epub 2020 Mar 23. PMID: 32200679; PMCID: PMC7176350.

Kato T, Tsukanaka A, Harada T, Kosaka M, Matsui N. Effect of hypercapnia on changes in blood pH, plasma lactate and ammonia due to exercise. Eur J Appl Physiol. 2005 Dec;95(5-6):400-8. doi: 10.1007/s00421-005-0046-z. Epub 2005 Sep 29. PMID: 16193339.

Larionova I, Kazakova E, Patysheva M, Kzhyshkowska J. Transcriptional, Epigenetic and Metabolic Programming of Tumor-Associated Macrophages. Cancers (Basel). 2020 May 29;12(6):1411. doi: 10.3390/cancers12061411. PMID: 32486098; PMCID: PMC7352439.

Li J, Hoskote A, Hickey C, Stephens D, Bohn D, Holtby H, Van Arsdell G, Redington AN, Adatia I. Effect of carbon dioxide on systemic oxygenation, oxygen consumption, and blood lactate levels after bidirectional superior cavopulmonary

anastomosis. Crit Care Med. 2005 May;33(5):984-9. doi: 10.1097/01.ccm.0000162665.08685.e2. PMID: 15891325.

Liu T, Melkus G, Ramsay T, Sheikh A, Laneuville O, Trudel G. Bone marrow adiposity modulation after long duration spaceflight in astronauts. Nat Commun. 2023 Aug 9;14(1):4799. doi: 10.1038/s41467-023-40572-8. PMID: 37558686; PMCID: PMC10412640.

Liu X, Li S, Cui Q, Guo B, Ding W, Liu J, Quan L, Li X, Xie P, Jin L, Sheng Y, Chen W, Wang K, Zeng F, Qiu Y, Liu C, Zhang Y, Lv F, Hu X, Xiao RP. Activation of GPR81 by lactate drives tumour-induced cachexia. Nat Metab. 2024 Apr;6(4):708-723. doi: 10.1038/s42255-024-01011-0. Epub 2024 Mar 18. Erratum in: Nat Metab. 2024 Dec 14. doi: 10.1038/s42255-024-01207-4. PMID: 38499763; PMCID: PMC11052724.

Manoharan I, Prasad PD, Thangaraju M, Manicassamy S. Lactate-Dependent Regulation of Immune Responses by Dendritic Cells and Macrophages. Front Immunol. 2021 Jul 29;12:691134. doi: 10.3389/fimmu.2021.691134. PMID: 34394085; PMCID: PMC8358770.

Mosienko V, Teschemacher AG, Kasparov S. Is L-lactate a novel signaling molecule in the brain? J Cereb Blood Flow Metab. 2015 Jul;35(7):1069-75. doi: 10.1038/jcbfm.2015.77. Epub 2015 Apr 29. PMID: 25920953; PMCID: PMC4640281.

Nechipurenko YD, Semyonov DA, Lavrinenko IA, Lagutkin DA, Generalov EA, Zaitceva AY, Matveeva OV, Yegorov YE. The Role of Acidosis in the Pathogenesis of Severe Forms of COVID-19. Biology (Basel). 2021 Aug 31;10(9):852. doi: 10.3390/biology10090852. PMID: 34571729; PMCID: PMC8469745.

Perciavalle V, Marchetta NS, Giustiniani S, Borbone C, Perciavalle V, Petralia MC, Buscemi A, Coco M. Attentive processes, blood lactate and CrossFit®. Phys Sportsmed. 2016 Nov;44(4):403-406. doi: 10.1080/00913847.2016.1222852. Epub 2016 Aug 24. PMID: 27556548.

Poliakov VV, Noskov VB. [Metabolic investigations in the 438-day space flight]. Aviakosm Ekolog Med. 2005 May-Jun;39(3):9-13. Russian. PMID: 16193920.

Rizzo A, Gimenez M, Horsky P, Saunier C. Influence d'une atmosphère de CO2 a 4% sur le comportement métabolique a l'exercice d'hommes jeunes [Metabolism during exercise in young men breathing 4% CO2 (author's transl)]. Bull Eur Physiopathol Respir. 1976 Jan-Feb;12(1):209-21. French. PMID: 1016774.

Rodgers GM, Fisher JW, George WJ. Lactate stimulation of renal cortical adenylate cyclase: a mechanism for erythropoietin production following cobalt treatment or hypoxia. J Pharmacol Exp Ther. 1974 Sep;190(3):542-50. PMID: 4370141.

Rodgers GM, Fisher JW, George WJ. The role of renal adenosine 3',5'-monophosphate in the control of erythropoietin production. Am J Med. 1975 Jan;58(1):31-8. doi: 10.1016/0002-9343(75)90530-6. PMID: 163577.

San-Millán I, Brooks GA. Reexamining cancer metabolism: lactate production for carcinogenesis could be the purpose and explanation of the Warburg Effect. Carcinogenesis. 2017 Feb 1;38(2):119-133. doi: 10.1093/carcin/bgw127. PMID: 27993896; PMCID: PMC5862360.

Sun Z, Han Y, Song S, Chen T, Han Y, Liu Y. Activation of GPR81 by lactate inhibits oscillatory shear stress-induced endothelial inflammation by activating the expression of KLF2. IUBMB Life. 2019 Dec;71(12):2010-2019. doi: 10.1002/iub.2151. Epub 2019 Aug 24. PMID: 31444899.

Terskov I.A., Gitelzon I.I., Kovrov B.G., et al. Closed system: man - higher plants. Novosibirsk. - Nauka, 1979, 160 p. Russian.

Trappe S, Costill D, Gallagher P, Creer A, Peters JR, Evans H, Riley DA, Fitts RH. Exercise in space: human skeletal muscle after 6 months aboard the International Space Station. J Appl Physiol. 2009b 106(4) :1159-68. doi: 10.1152/japplphysiol.91578.2008.

Vlasova N.V., Gitelzon I.I., Okladnikov Y.N. Lipid metabolism in man under nutrition with lyophysized diet during 4-6 months in conditions of closed system of life support. Voprosy Pitaniya, 1976, No. 2 p. 17-20. Russian

Warburg O. On the origin of cancer cells. Science. 1956 Feb 24;123(3191):309-14. doi: 10.1126/science.123.3191.309. PMID: 13298683.

Wasserman K, Mcilroy MB. Detecting the threshold of anaerobic metabolism in cardiac patients during exercise. Am J Cardiol. 1964 Dec;14:844-52. doi: 10.1016/0002-9149(64)90012-8. PMID: 14232808.

Wasserman K, Van Kessel AL, Burton GG. Interaction of physiological mechanisms during exercise. J Appl Physiol. 1967 Jan;22(1):71-85. doi: 10.1152/jappl.1967.22.1.71. PMID: 6017656.

Wu P, Zhu T, Huang Y, Fang Z, Luo F. Current understanding of the contribution of lactate to the cardiovascular system and its therapeutic relevance. Front Endocrinol (Lausanne). 2023 Jun 15;14:1205442. doi: 10.3389/fendo.2023.1205442. PMID: 37396168; PMCID: PMC10309561.

Yeung DK, Griffith JF, Antonio GE, Lee FK, Woo J, Leung PC. Osteoporosis is associated with increased marrow fat content and decreased marrow fat unsaturation: a proton MR spectroscopy study. J Magn Reson Imaging. 2005 Aug;22(2):279-85. doi: 10.1002/jmri.20367. PMID: 16028245.

Zhang D, Tang Z, Huang H, Zhou G, Cui C, Weng Y, Liu W, Kim S, Lee S, Perez-Neut M, Ding J, Czyz D, Hu R, Ye Z, He M, Zheng YG, Shuman HA, Dai L, Ren B, Roeder RG, Becker L, Zhao Y. Metabolic regulation of gene expression by histone lactylation. Nature. 2019 Oct;574(7779):575-580. doi: 10.1038/s41586-019-1678-1. Epub 2019 Oct 23. PMID: 31645732; PMCID: PMC6818755.